\documentclass[conference,anonymous]{IEEEtran}
\IEEEoverridecommandlockouts
\usepackage{cite}
\usepackage{amsmath,amssymb,amsfonts}
\usepackage{algorithmic}
\usepackage{graphicx}
\usepackage{textcomp}
\usepackage{xcolor}
\usepackage{braket}
\usepackage{subcaption}
\usepackage{comment}

\def\BibTeX{{\rm B\kern-.05em{\sc i\kern-.025em b}\kern-.08em
    T\kern-.1667em\lower.7ex\hbox{E}\kern-.125emX}}

\begin{document}
\graphicspath{ {./} }
\title{Just-in-time Quantum Circuit Transpilation Reduces Noise
}

\author{\IEEEauthorblockN{Ellis Wilson, Sudhakar Singh, Frank Mueller}
\IEEEauthorblockA{\textsuperscript{1}\textit{North Carolina State University}, \{ejwilso2,ssingh46\}@ncsu.edu, mueller@cs.ncsu.edu}
}

\maketitle

\begin{abstract}
Running quantum programs is fraught with challenges on on today's
noisy intermediate scale quantum (NISQ) devices. Many of these
challenges originate from the error characteristics that stem from
rapid decoherence and noise during measurement, qubit connections,
crosstalk, the qubits themselves, and transformations of qubit state
via gates. Not only are qubits not ``created equal'', but their noise
level also changes over time.  IBM is said to calibrate their quantum systems
once per day and reports noise levels (errors) at the time of such
calibration.  This information is subsequently used to map circuits to
higher quality qubits and connections up to the next calibration
point.

This work provides evidence that there is room for improvement over
this daily calibration cycle. It contributes a technique to measure
noise levels (errors) related to qubits immediately before executing
one or more sensitive circuits and shows that just-in-time noise
measurements benefit late physical qubit mappings. With this just-in-time
recalibrated transpilation, the fidelity of results is improved over
IBM's default mappings, which only uses their daily calibrations. The
framework assess two major sources of noise, namely readout errors
(measurement errors) and two-qubit gate/connection errors. Experiments
indicate that the accuracy of circuit results
improves by 3-304\% on average and up to 400\% with on-the-fly circuit
mappings based on error measurements just prior to application execution.

\end{abstract}

\begin{IEEEkeywords}
quantum computing, errors, dynamic compilation
\end{IEEEkeywords}

\section{Introduction}

Today's quantum computing devices and those of the foreseeable future
are referred to as Noisy Intermediate Scale Quantum (NISQ) computers due
to the noise inherent in the systems and the small number of quantum
bits (qubits) available for calculations~\cite{murali2019noise}.
Even when calculations can
be performed with a small number of qubits, the noise in the quantum
systems frequently produces incorrect results, which presents a
challenge in using quantum computation. Consequently, techniques to
identify, mitigate, and tolerate noise and even errors in calculations
are of considerable importance for quantum computation amid this noisy
reality.
	
Different types of errors can be distinguished. The most commonly
reported errors are:
\begin{itemize}

\item Readout errors: These are errors in interpreting
the state of the qubit at the end of the calculation, e.g., reading a
qubit in the $\ket{0}$ state as being in the $\ket{1}$ state.

\item Single qubit gate errors: These occur when applying gates to
  a single qubit causes small changes in the qubit state, which can
  accumulate over deep circuits with long sequences of gates.

\item Two-qubit gate errors: These result from interaction between two
  qubits under a two-qubit gate operation (e.g., both qubits of a CNOT
  gate);

\item Decoherence errors: These are due to the decay of state over
  time in today's quantum devices, and they are referred
  to as T1, T2, and T2* --- but are not addressed in this work.

\item Cross-talk errors: These result when the state of a qubit or a
  resonator between qubits influences the state of another qubit or
  resonator in close vicinity --- but are again beyond the scope of
  this work.

\end{itemize}

Each of these errors can vary from one qubit to another and also from
one connection (coupling) to another; some qubits/connections
experience less noise and fewer errors than others. To make matters
more complicated, the qubits themselves change over time in an
unpredictable fashion (due to the quantum nature of the qubit system),
leading to a need to re-calibrate the qubits and recalculate these
errors, e.g., once per day on IBM Q systems.
	
One way to reduce errors in quantum computations, especially on
systems with more physical qubits than necessary for the circuit in
question, is to try to map the circuit onto the most appropriate
qubits during a process called ``transpilation''. Transpilation
traditionally considers the mapping of logical qubits in a program
onto a physical NISQ device with limited qubit connectivity and native
gate operations. This may require a high-level gate (e.g., X/Y/Z
rotation) to be translated into one or more low-level gates (e.g.,
U1/U2/U3 for IBM) with specific phase angles. Transpilation may result
in logical qubits being moved (via swap operations) from one physical
qubit to another throughout a circuit during its execution. More
contemporary transpilation considers virtual-to-physical mappings to
the highest fidelity qubits and connection between qubits to reduce
the overall error~\cite{Tannu:2019:QCE:3297858.3304007,murali2019noise}.
These optimizations are clearly non-trivial, as
many mappings exist in this multi-dimensional non-linear optimization
space. For example, the highest fidelity qubits for one circuit may
not provide the connections for two-qubit gates of another circuit.
It is therefore important to have accurate fidelity data of the
physical machine for which a circuit is being transpiled. Different
transpilers exist, each of which accept different types of
statistical error values per qubit and per coupling between qubits before
attempting to provide a high quality mapping. For IBM's quantum
computers, these error metrics are derived from calibration runs of
circuits that measure qubits and compare values with reference
results. Such calibration occurs usually once per day, and error
metrics are published on IBM's websites and can also be obtained from
the Qiskit API~\cite{Qiskit} for the latest calibration run.

We have performed a series of experiments that, after an initial
stable phase, uncovered a quick deterioration in fidelity of qubit
gates, measurements and couplings not too long after calibration.
These experiments included micro-benchmarks to (a) prepare an $n$
qubit circuit with initial state $\ket{0}^n$ followed by simply
measuring each qubit and (b) subjecting each qubit to a series of X
(NOT) gates before measurement. When repeated hourly, no clear trend
became visible.  Neither could we detect when the original calibration
took place, nor did we observe a gradual de-calibration.  
When employing longer and more complex circuits and directly comparing
different error values, we found that while some qubits remained
stable, other qubits showed significant variations in fidelity
throughout the day.
	
These findings motivated us to experiment with obtaining calibration
data ourselves, use them in just-in-time transpilation, and then
observe errors for this transpiled circuit compared to one
transpiled with IBM's calibration data. Clearly, if
virtual-to-physical mappings differ between just-in-time transpilation
vs. default transpilation, the fidelity of results can be expected to
differ as well.

We chose to focus the investigation on readout errors and two-qubit
gate errors in particular, the most significant errors in
magnitude. This is also motivated by readout errors affecting every
circuit and two-qubit gate errors consistently being about an order of
magnitude higher than single qubit gate errors, i.e., the qubit
placement of two-qubit gates during transpilation is of high
importance for overall fidelity.

Readout errors were determined by immediately measuring a newly
prepared qubit and subsequently applying a single X (NOT) gate before
measuring the same qubit again. We compared the results to the
expected values (of all $\ket{0}$ or all $\ket{1}$, respectively) to
obtain the percent error.
Two-qubit gate errors were determined by utilizing
IBM's built-in randomized benchmarking capability to obtain an
error value.
We then subjected transpilation to our error values instead of the
default IBM ones (from daily calibration). This resulted in different
qubit mappings leading to an improvement of 3-304\% on average, and up to 400\%.

\section{Design}
\label{sec:design}

The design of our just-in-time transpilation was driven by an initial
experiment followed by a methodology to address shortcomings of the
current system. While observations are specific to IBM Q devices, the
methodological approach is more generic and may transfer to other NISQ
devices.

\subsection{Motivating Experiments}

Our first objective was to determine whether or not the fidelity of
qubits varied significantly throughout the day.
If the fidelity of qubits did not vary significantly between two
calibration instances, any effort to repeatedly assess the error rates
would likely not contribute to fidelity improvements.
To test our hypothesis of variations, we conducted experiments with a
number of circuits assessing reported errors throughout the day.

The main focus centered on the qualitative aspect of qubit change, i.e.,
do qubits provide different results in fidelity over time between
calibrations, rather than absolute errors. To this end, experiments 
were limited to simple circuits to assess readout errors or errors
due to successive Pauli gates, which were repeated every hour.

\begin{figure}
  \includegraphics[width=\linewidth]{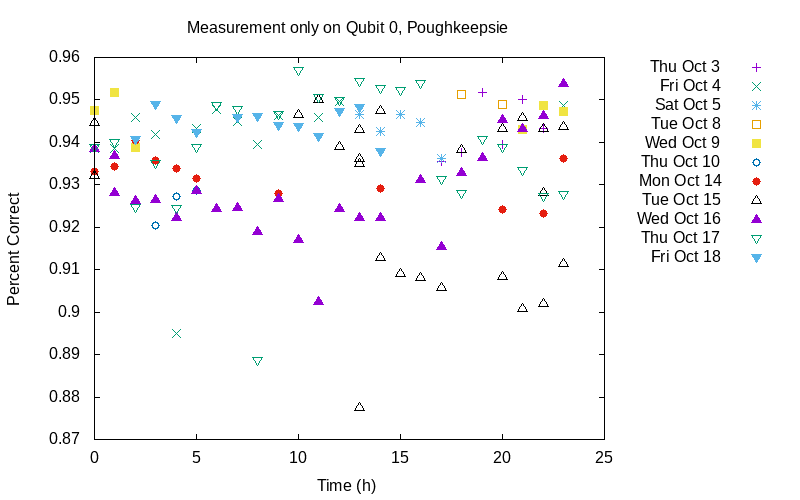}
  \caption{Measurements with no gates on qubit 0 of IBM Poughkeepsie
    over time. The fidelity of readouts for the qubit varies in a
    chaotic (non-predictable) manner. Results for other qubits and
    $\ket{1}$  circuits are similar, figures omitted due to space.}
  \label{fig:rd0PoughQ0}
\end{figure}

The first experiment focused on readout errors without any gates, where
a qubit was initialized ($\ket{0}$ state) and then immediately measured. 
The second experiment assessed readouts for a qubit after a Pauli X
(NOT) gate., i.e., the $\ket{1}$ state.
Fig.~\ref{fig:rd0PoughQ0} depicts hourly measurements (x-axis) over
the percentage of correct results (y-axis) on different days (colored
data series) in 2019 on the IBM \textit{Poughkeepsie} device (20 qubits).
The results show that qubits do not remain stable between
calibrations. This behavior was observed across qubits and different
IBM Q devices.

While readout results of these circuits usually did not change
drastically from hour to hour, they did change in an unpredictable
manner, sometimes resulting in better accuracy, sometimes in worse.
This made it impossible to infer or reverse engineer when the
calibration actually took place, i.e, we did not observe a drastic
change in quality for qubits when measured. We also tested circuits
with many gates in a less rigorous manner and observed similar
results.
Based on prior work, we know that circuits with virtual qubits
were mapped onto physical ones via transpilation (at optimization
level 3) taking IBM's error data from the last calibration into
account~\cite{murali2019noise,Tannu:2019:QCE:3297858.3304007}.
This led us to the new hypothesis that, when selecting physical qubits
to which circuits are to be mapped, a new set of error measurements
for just-in-time transpilation might improve the overall fidelity.

\subsection{Error Selection}

A number of different types of errors are taken into account when
mapping circuits to qubits, where some of these errors are more
prevalent than others as indicated by the respective error metrics.
For example, T1 and T2 errors are significant in long circuits but not
in short ones. 
Gate errors will be present in all circuits, but more so in long
circuits using many gates.
Readout errors need to be taken into account in all circuits. 
If some errors are more significant than others, those errors dominate
the mapping decisions, while other, less significant ones will only
marginally contribute.

We decided to focus on two types of errors, those from readouts
and those from two-qubit gates.
Readout errors affect any circuit, and their probability is relatively
high on today's NISQ computers.
Readout errors are reported to be in the order of $10^{-2}$ for IBM Q devices.
We even observed that sometimes they can be as high as 10\%. 

We also focus on two-qubit gate errors for the same reason: They have
an equally high error rate (both reported by IBM and observed by us).
In contrast, single-qubit gate errors are reported to be lower ($10^{-3}$
for IBM Q devices), and they were also an order of magnitude smaller
than readout or two-qubit errors in our experiments.

\subsection{Methodological Error Collection}

The challenge at hand is to reliably collect error characteristics of
a physical quantum device that can subsequently be used to map
circuits to physical qubits such that overall fidelity can be increased.
Readout errors are the easiest to be measured, simply by
constructing a circuit that minimizes any of the other types of errors
while producing a known measurement value.
To minimize gate and time-based errors, the qubits are measured as
quickly as possible and with the fewest number of gates. 
We observe that reading $\ket{0}$ and $\ket{1}$ states each have
different error
rates~\cite{DBLP:conf/micro/TannuQ19,DBLP:conf/micro/TannuQ19a}. Hence,
we utilize two circuits per qubit to characterize readout errors. The
first circuit prepares a qubit in the $\ket{0}$ state (as quickly as
possible) and measures it, while the second one prepares the qubit in
the $\ket{1}$ state via a single X (NOT) gate before measuring.

Gate errors present more of a challenge to be assessed. Recall that we focus
on two-qubit gates here due to their higher error rates compared to single qubit
gates.
Two-qubit gate errors are determined for each pair of connected
qubits that can be captured through randomized benchmarking, which
uses randomized sequences of gates of increasing length
resulting in a known $\ket{0}$ state on qubits. By comparing actual
measurements to this known value, error rates are determined.
This is described in more detail in~\cite{Magesan_2011}.

These error characteristics are subsequently used for just-in-time
transpilation of circuits for mapping to physical qubits with high
fidelity for couplings/connections within the circuit and high
measurement quality of selected qubits.

\section{Implementation}
\label{sec:impl}

We decided to implement our high-level design of just-in-time
transpilation for IBM Q devices using Qiskit. This involves data
collection on errors on an IBM Q device, subsequent transpilation of
benchmarks via Qiskit at an optimization level that takes errors into
account when mapping to physical qubits, and running these benchmarks
on the same IBM Q device.

In order to test whether just-in-time error measurement improves
performance over using the daily calibrations, we need to to reliably
collect data on errors and, for a fair comparison, in a similar manner
to how IBM collects data and reports errors during their daily
calibrations.
	
Due to the nature of IBM’s qubits (and other technologies as well),
the error for reading a qubit in the $\ket{0}$ ground state state is
much lower than reading a qubit in the $\ket{1}$ excited state, which
is less stable~\cite{DBLP:conf/micro/TannuQ19,DBLP:conf/micro/TannuQ19a}.
IBM determines readout error rates for each state as well as the
average of both, which it reports as the readout error.
These errors are relatively easily obtained. As described in our
motivating experiments, to assess errors for readouts of the $\ket{0}$
state one merely needs to measure immediately after preparing a qubit.
Similarly, the $\ket{1}$ state is read out after a qubit is prepared
and subjected to a single X (NOT) gate.
The observed level of error between the single qubit gate and the
readout error, especially when in the $\ket{1}$ state, shows that the
contribution of the X gate to the error is negligible (about an order
of magnitude lower than the readout error).  The readout error is this
calculated as the percent of incorrect results returned from the
respective circuits.
	
The two-qubit error requires more complex circuits. We employ Qiskit's
randomized benchmarking capabilities, which can automate the process
of data collection. 
These randomized benchmarks consist of circuits with two qubits that
are generated such that their output is an ``Error per Clifford''
value, which is proportional to the two-qubit error itself.

Obtaining these error metrics for each qubit is a computationally
intensive task. Due to limited compute cycles, we decided to combine
many of the individual qubit measurements into a single multi-qubit
measurement. While this ignores the impact of
qubit crosstalk, it still remains useful as any circuit, including our
benchmarks, also utilizes multiple qubits, often in close physical
vicinity to reduce the number of swaps in transpiled programs.
We split the two-qubit gate errors up such that only one coupler of a given
qubit was assessed in terms of error at a time.
On the IBM Q devices used here, the maximum degree of a qubit is three,
i.e., we ran a total of three jobs to capture all two-qubit errors.
Once each of these errors had been measured, we assessed the
virtual-to-physical mappings. This allowed us to report errors
for each physical qubit.

As we are focusing on IBM Q devices, we decided to leverage Qiskit's
built in transpiler using the highest optimization level available
(level 3).  When compiling benchmarks (or other application programs),
this level-3 transpilation triggers an optimization for
virtual-to-physical mappings~\cite{murali2019noise}.

\section{Experimental Setup}
\label{sec:framework}

We conducted experiments on various IBM Q devices throughout different
days and different times as well as repeatedly during a
particular day. In every experiment, we first manually measured the
CNOT and readout errors and then, based on this error information,
transpiled our circuits before sending them to the devices to execute
them. We kept track of the circuit layouts post-transpilation and
their performance with respect to accuracy. Next, we describe the
individual aspects of this setup.

\subsection{Device Information}

\begin{figure}[htb]
  \begin{subfigure}{\columnwidth}
    \includegraphics[width=\linewidth]{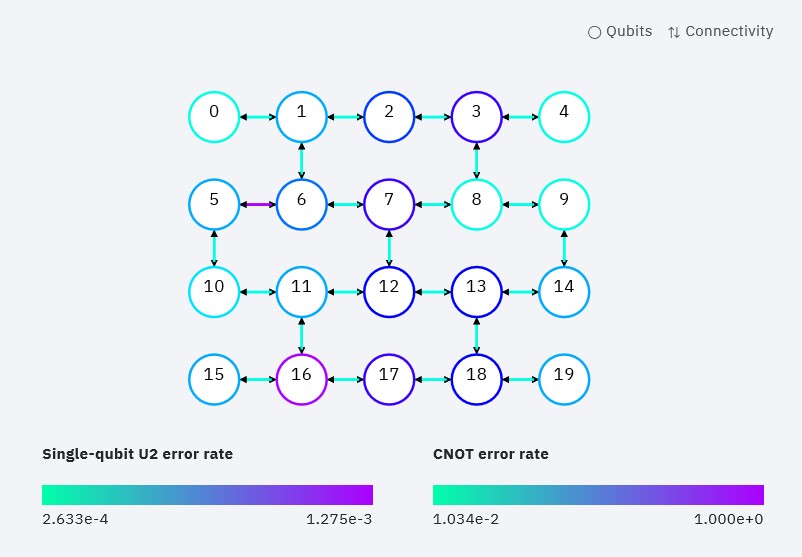}
    \caption{IBM Q Almaden device}
  \end{subfigure}
  \begin{subfigure}{\columnwidth}
    \includegraphics[width=\linewidth]{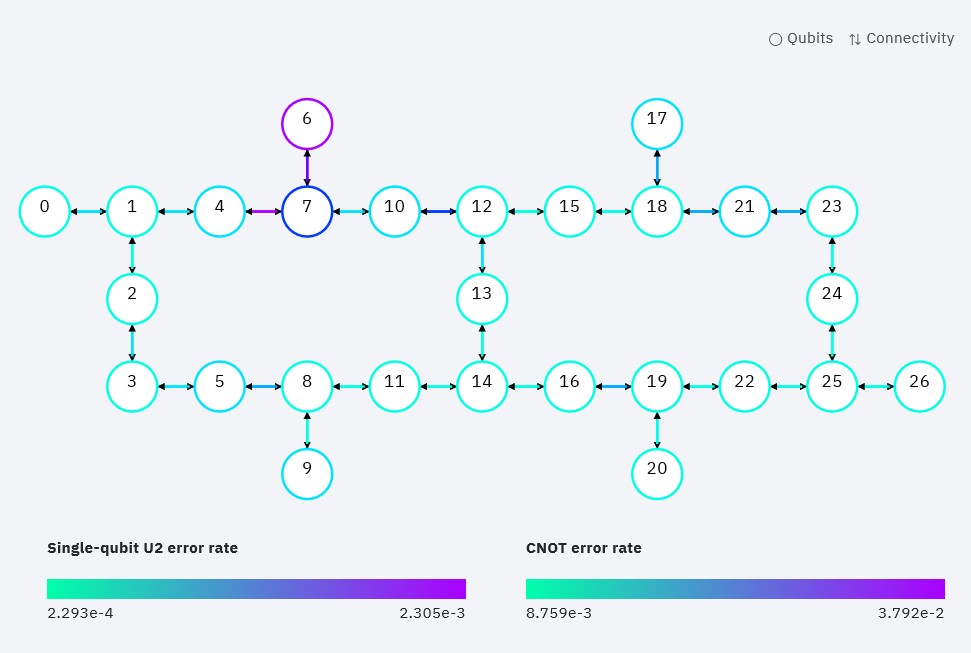}
    \caption{IBM Q Paris device}
  \end{subfigure}
\caption{Qubit connectivity with colored mappings for qubits and
  connections corresponding to COTD information on a heatmap
  range. Source:~\cite{ibmqexp}. }
\label{fig:ibmq_backends}
\end{figure}

We performed our experiments primarily on two IBM Q devices,
\textit{Almaden} (20 qubits) and \textit{Paris} (27 qubits). The
rationale was to select backends with a sufficient number of possible
virtual-to-physical qubit mappings so that the transpilation procedure
could adapt mappings to error data. Both devices allow a total of 900
circuits to be sent in one job. Availability of these devices
presents another challenge, as they tend to be busy with many jobs in
the queue, which meant that the calibration job was running an hour or
more before the benchmark jobs as the latter can only be submitted
after transpilation taking errors from the former job into account.
This assesses just-in-time transpilation in a normal user scenario
with so-called ``fairshare'' queuing. In addition, we conducted
experiments in ``dedicated'' mode, available only to select users,
where calibration and benchmark jobs can be run within minutes of
each other, again after transpilation of benchmarks based on error
data from immediately preceding calibration.
%
Figure~\ref{fig:ibmq_backends} depicts the physical qubit topology of the two
backends, Almaden and Paris, with a snapshot of the calibration-of-the-day (COTD) data
encoded as colors according to the respective heatmap of the device.

\subsection{Benchmarking Circuits}

We selected a number of circuits for just-in-time transpilation also
used in prior work~\cite{murali2019noise}. The characteristics of the
selected benchmarks were based on the ability to scale single qubit
gates, two-qubit gates, circuit depth and circuit width (i.e., the
number of qubits).
These benchmarks can be parametrized by the number of qubits, $n$, and
are:


\begin{enumerate}
\item bv(n): the Bernstein-Vazirani algorithm that learns an $n$-bit string
  encoded in a function and reads out $n+1$ qubits;
\item hs(n): the n-bit/qubit hidden-shift algorithm that determines
  the constant by which the input of one function is increased
  (shifted) relative to that of another function, where $n$ qubits are
  measured;
\item qft(n): the n-bit/qubit quantum Fourier transform algorithm,
  which is used in many other quantum algorithms as a building block
  with $n$ qubits measured;
\item toffoli(n): the n-qubit ``universal'' Toffoli gate that can be
  specialized for a number of arithmetic operations depending on
  parameters, $n+1$ qubits are read out;
\item adder(n): an n-bit adder algorithm using $2 \times n + 2$ qubits
  and $n+1$ readouts.
\end{enumerate}

Algorithms 1-3 include Hadamard gates and conditional rotational
gates, yet still have known reference
outputs. Conversely, algorithms 4-5 consist of Pauli gates, C-NOT
gates or CC-NOT gates (with two conditionals), the latter of which can
be transpiled into a sequence of single qubit (Hadamard and
rotational) gates and six C-NOT gates, again with known expected
outputs.

\subsection{Qiskit Experiments}

\begin{figure}
\includegraphics[width=\linewidth]{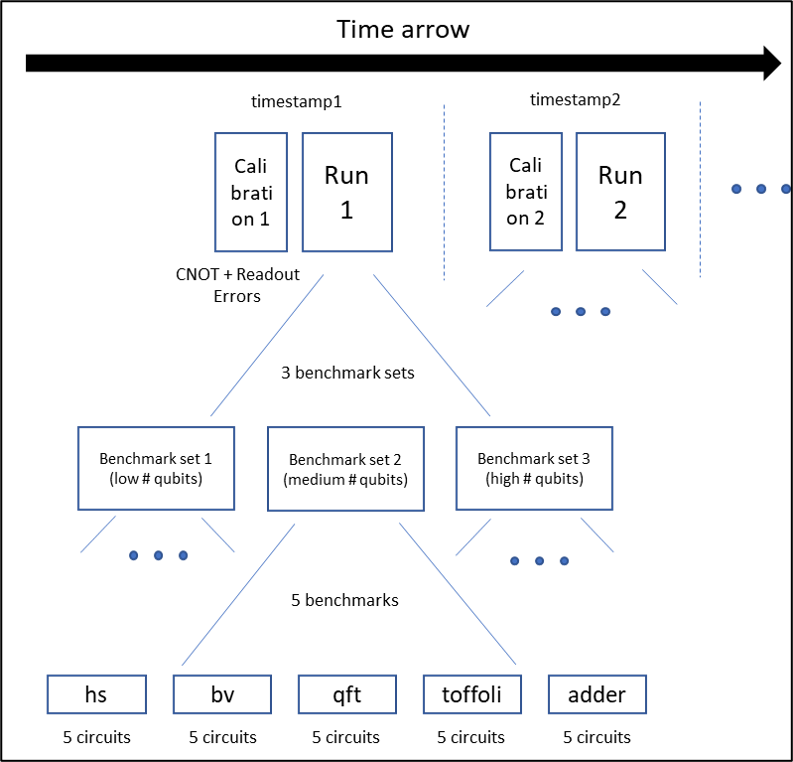}
\caption{Calibration and benchmark circuits as arranged in our job framework}
\label{fig:bench_frame}
\end{figure}
Qiskit provides an interface for sending multiple quantum circuit
experiments to the device in a single job. The maximum number of these
experiments depends on the type of device. As an example, 
the \textit{Almaden} device accepts a total of 900 circuits in a
single job. CNOT and readout error
calibrations are performed in our experiments using the calibration
circuits detailed in previous section. These are run repeatedly at a
particular time of the day, as shown in Figure~\ref{fig:bench_frame}
(timestamp1, timestamp2 etc.). With the resulting error data, several
benchmark circuits are transpiled. We investigate 5 circuits
representing the above benchmark codes per run, where each circuit is
executed for 4096 shots, i.e., repeated circuit executions with a
measurement. Further, we have 3 sets of these benchmarks with
increasing number of qubits as shown in the figure (low/medium/high
number of qubits) in a single job. In total, a single benchmark measurement job contains 75 circuits, i.e., 25
circuits per benchmark set and 5 circuits for each individual
benchmark with exactly the same circuit and mapping since they are transpiled
together with the same calibration data using the
\textsl{qiskit.compiler.transpile} function. We execute several
circuits for a particular benchmark, each with 4096 shots, as we
observed that for certain qubits and connections significant
variations in the accuracy exist across different circuit executions
within the same job. In summary, each benchmark job that gets
submitted to the device is just-in-time transpiled with the latest
error data obtained by our measurements --- instead
of the default COTD data from IBM. Depending on
the experiment, these jobs are either run at different times on
different days in ``fairshare'' queuing, or they are repeatedly run
throughout the day in ``dedicated'' time slots to capture the variance
in the accuracy of benchmark circuits. Notice that dedicated execution
is a novel feature that became available only in late May 2020.

\section{Results}
\label{sec:results}

We first report results in the default user mode, followed by dedicated mode.
We then perform a sensitivity analysis with respect to circuit layouts
before discussing overall findings and implications.

\subsection{Fairshare User Mode}

These experiments on IBM Q \textit{Almaden} consist of a first job running all benchmarks
resulting from level 3 transpilation using IBM's error data, followed
by eight instances of two jobs, one for measurement to obtain
refreshed error data and a second to run all benchmarks transpiled at
level 3 with the fresh error data. Percent accuracy relative to
expected results (y-axis) is plotted for each benchmark run (x-axis).

\begin{figure}[t!]
  \includegraphics[width=\linewidth]{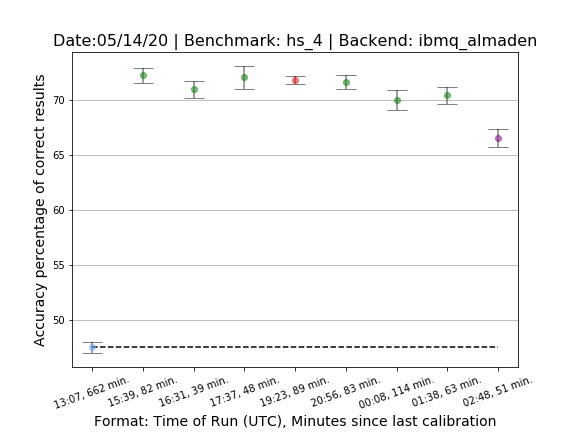}
  \includegraphics[width=\linewidth]{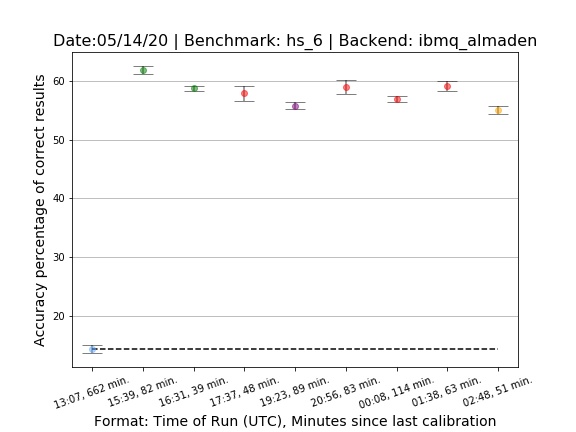}
  \includegraphics[width=\linewidth]{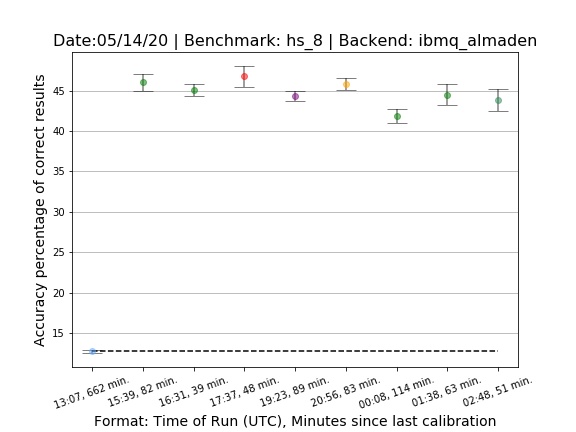}
  \caption{Accuracy of Hidden Shift for 4/6/8 qubits
    (upper/middle/lower graphs) at different times during the day with
    prior calibration in minutes.}
  \label{fig:hs_almaden_5_14}
\end{figure}

Figure~\ref{fig:hs_almaden_5_14} depicts results for Hidden Shift (hs) with
4, 6, and 8 qubits, where the x-axis indicates the
time (during the day) when the benchmark run started and the number of
minutes prior to which error data was measured.

For hs(4), the top graph in the figure, the first data point shows an
accuracy of 47\% (with a small standard deviation indicated by the
whiskers) for IBM's reference calibration 10 hours earlier. This is
the reference run (dashed line) for our experiments. The remaining
data points are showing 67-73\% accuracy for our runs, spaced in 1-2
hour intervals whenever the job queue scheduled runs, with prior error
data obtained 39 minutes to nearly 2 hours earlier. The different
colors of data points indicate distinct layouts of virtual-to-physical
qubits. Our layouts differ from IBM's layout due to the refreshed
error data, which provides the benefits in accuracy.

For hs(6) and hs(8) in middle and lower graphs of Figure
\ref{fig:hs_almaden_5_14}, our results
have an even higher improvement in accuracy over IBM's reference
layout, with our layouts changing from hour to hour. Overall, IBM's
accuracy is reduced from 47\% to 27\% to 12\% for hs(4), hs(6) and
hs(8), respectively. This reflects the higher number of qubits used
and longer depth of a given circuit.  With our just-in-time
transpilation, the values are much higher: 70\%, 58\%, and
45\% on an average for hs(4), hs(6) and hs(8), respectively.

Results for other benchmarks are similar in trend, albeit with
different absolute accuracies/improvements with figures omitted
due to space.
Relative improvement in accuracy ranges from 8-48\% for bv, 48-304\%
for hs, 45-69\% for qft, 133-155\% for toffoli, and 12-42\% for adder,
with maximum improvements sometimes as high as 400\%, i.e., a factor
of four improvement in accuracy.  We also observe that toffoli, qft
and adder have a higher standard deviation.

\textbf{Observation 1:} Just-in-time transpilation tends to improve
the relative accuracy of measured results on average by 8\%-304\% and up to
400\% in extreme cases in fairshare user mode. Best layouts change at least hourly.

\begin{figure}[htb]
  \includegraphics[width=\linewidth]{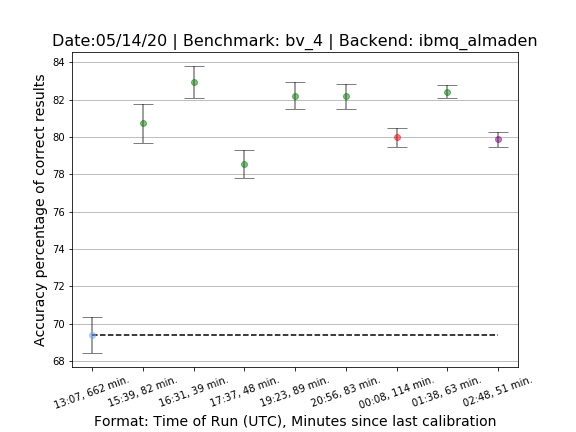}
  \includegraphics[width=\linewidth]{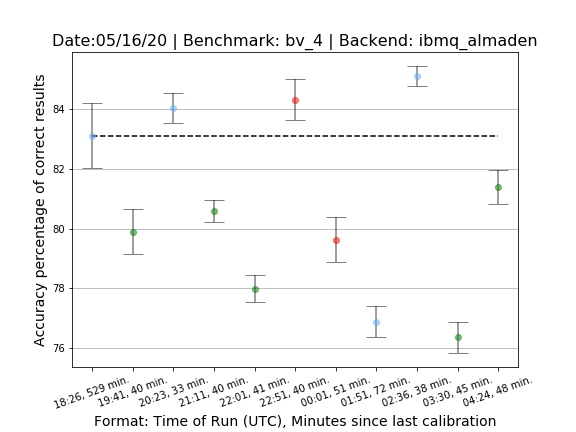}
  \caption{Accuracy of Bernstein-Vazirani with 4 Qubits on 5/14/20 (upper)
    and 5/16/20 (lower)}
  \label{fig:bv_almaden_14_16}
\end{figure}

Figure~\ref{fig:bv_almaden_14_16} depicts results for
Bernstein-Vazirani (bv) with 4 qubits on two different days.  On the
first day (upper graph), trends are similar to hidden-shift, where the
accuracy of just-in-time transpiled benchmarks
throughout the day is consistently higher (around 82\%) than those
transpiled with using IBM's COTD (69.5\%). The difference between our
measurements is relatively small (+/-5\%).
But on a different day (lower figure), results are mixed as the
benchmarks transpiled with IBM's COTD show higher accuracy (83\%)
while many of our just-in-time transpilations result in lower accuracy
(as little as 76\%) while others are slightly better (up to 85\%) than
IBM's reference. Interestingly, all the benchmarks show more significant
standard deviations (wider whiskers) in the lower graph, even though
IBM's calibration was about 10 hours prior in both cases.
Closer inspection reveals that the same IBM layout mapping (blue dot) also
provides slightly better results (3rd and 9th data point), yet worse
results at a different time (8th data point).

\textbf{Observation 2:} Benefits of just-in-time transpilation vary
from day to day, even for the same layouts of qubits on a device.

While we observe such variation, we actually cannot provide absolute
conclusions from this data as we only ran benchmarks with IBM's COTD
layout once, and only hours apart from our just-in-time
experiments. This led us to conduct a set of experiments in dedicated
mode close together in time, once this mode became available. This is
discussed in the next subsection.

The QFT circuit (figures omitted due to space) contains a large number
of two-qubit gates and thus results in lower overall accuracy and also
declining accuracy as circuits are scaled up from 4 over 6 to 8 qubits.
As before, IBM's accuracy is generally lower than ours (30\% vs. 18\%
for 4 qubits) but the total value becomes unreasonably low for 8
qubits (IBM: 0.85\%, ours: 1.3\%), even though our results are still
better on one of the days.
However, on another day, only half of our just-in-time calibrations
resulted in benefits over IBM's, still with the same low accuracy
under qubit scaling.
%
%
The Toffoli and adder benchmarks show trends similar to the QFT benchmark.

\textbf{Observation 3:} As the number of qubits is scaled up, total
accuracy drops significantly to the point where few results remain
correct, even with just-in-time transpilation. IBM's results
remain inferior to our just-in-time method.

\subsection{Dedicated Mode}

In regular user mode, fairshare queuing \cite{Fairshare} on IBM Q devices prevents a
calibration job to be run back-to-back with benchmarks as just-in-time
transpilation requires the error data from the calibration run, and
typical queue delay is on the order of hours for IBM Q Hub devices (or
even days for public devices). While we showed that qubit fidelity in
terms of readout and coupling errors varies, our prior results were
inconclusive with respect to the rate at which these variations take
place.

A novel dedicated queuing mode allows the reservation of time slots of
fixed lengths at a given time of the day. This allows us to reserve a
slot long enough to run a calibration test to obtain readout and
coupling errors, run benchmarks using IBM's error data while
transpiling with our newly obtained error data, and then run the
just-in-time transpiled benchmarks based on our error data. These
three jobs run back-to-back within 15 minutes. This experiment was
repeated 8 times during a 24-hour period.
Dedicated queuing was available on IBM's \textit{Paris} device with 27 qubits.

\begin{figure}
  \includegraphics[width=\columnwidth]{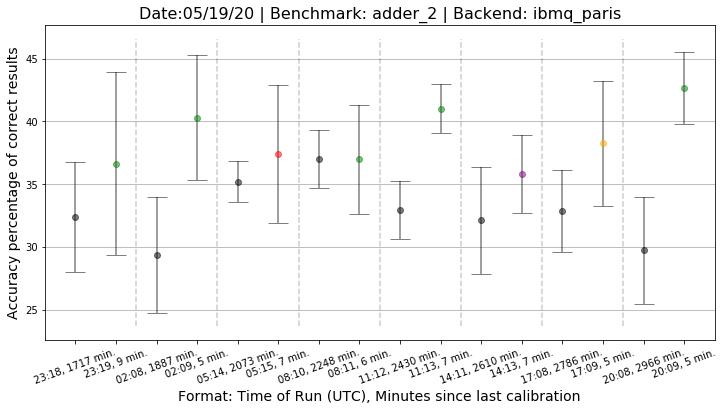}
  \includegraphics[width=\columnwidth]{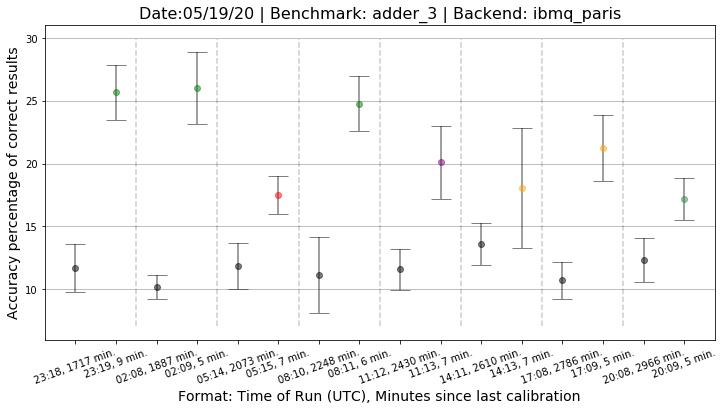}
  \includegraphics[width=\columnwidth]{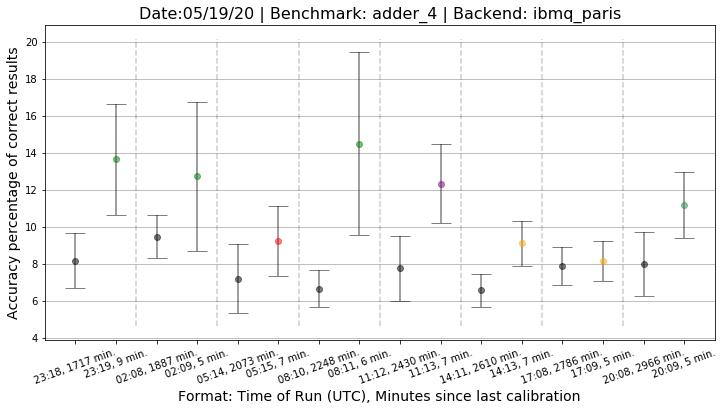}
  \caption{Accuracy of Adder for 4/6/8 qubits
    (upper/ middle/ lower graphs) at different times during the day with
    prior calibration in minutes.}
  \label{fig:adder_paris_5_19}
\end{figure}

Figure~\ref{fig:adder_paris_5_19} depicts the accuracy for a 2+2, 3+3
and a 4+4 adder (upper/middle/lower graphs) in dedicated mode. Black
dots indicate IBM's layout based on their COTD errors obtained 28-49
hours earlier. Notice that the device was {\em not} recalibrated
during this period, which indicates that IBM even calibrates less
frequently than the 24 hours that are commonly cited. Each set of
(black, colored) data points runs within the same time slot and should
be related to one another in comparisons.

For the first graph, we observe that IBM runs (black) vary
significantly in accuracy over time, as much as 29-37\%, i.e., a given
calibration with COTD error data does not provide consistent results.
We further observe that when any IBM run (black) is followed by our
just-in-time transpiled run (colored) minutes later, the latter
always provides higher accuracy. Standard variations are sometimes
higher, sometimes lower with no clear pattern. As circuit sizes are
scaled up (middle/lower graphs), this trend still holds, even as
absolute accuracy becomes smaller due to wider and deeper
circuits. The benefits of just-in-time transpilation are more
pronounced in the 3+3 adder (middle graph), without any clear cause as
these three benchmarks ran back-to-back (cf. absolute times indicated
on the x-axis). Just-in-time transpilation always resulted in a
different circuit than IBM's default transpilation, and the former
resulted in notable savings --- with the one exception of adder(4) in
the 2nd to last pair of (black, yellow) dots, where our benefit is
smaller. Layouts change between hourly slots.
These results generalize to other benchmarks with higher (bv, hs) or
lower (qft, toffoli) absolute savings. We did see occasional outliers as
discussed in the next subsection.
We summarize these findings as the following observation.
\begin{figure*}
  \centering
  \begin{subfigure}{.3\textwidth}
    \includegraphics[width=\textwidth]{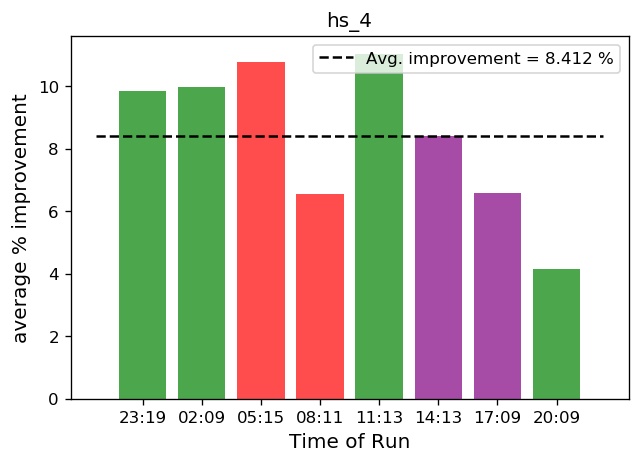}
  \end{subfigure}
  \hfil
  \begin{subfigure}{.3\textwidth}
    \includegraphics[width=\textwidth]{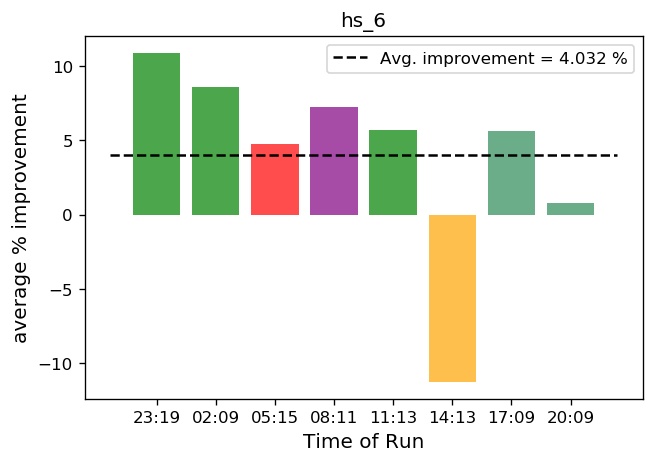}
  \end{subfigure}
  \hfil
  \begin{subfigure}{.3\textwidth}
    \includegraphics[width=\textwidth]{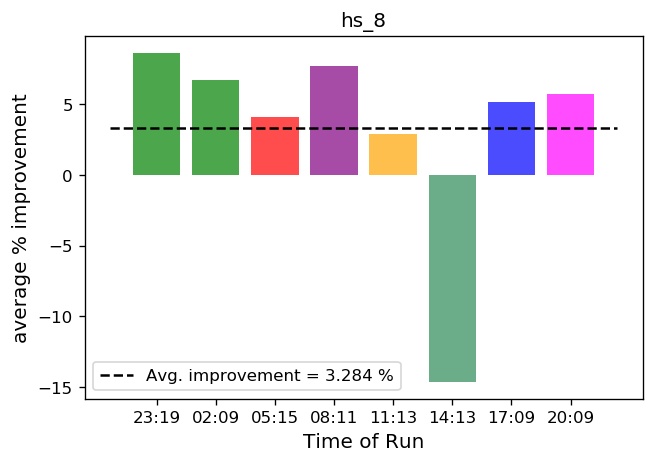}
  \end{subfigure}

  \begin{subfigure}{.3\textwidth}
    \includegraphics[width=\textwidth]{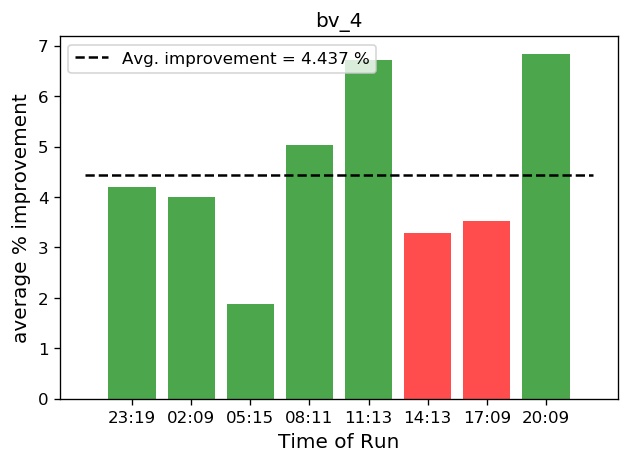}
  \end{subfigure}
  \hfil
  \begin{subfigure}{.3\textwidth}
    \includegraphics[width=\textwidth]{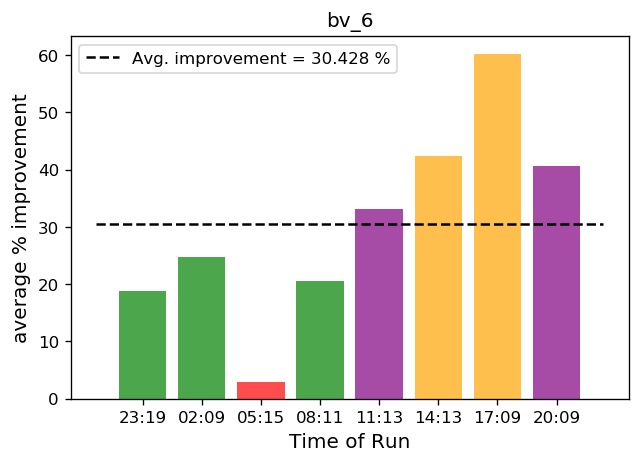}
  \end{subfigure}
  \hfil
  \begin{subfigure}{.3\textwidth}
    \includegraphics[width=\textwidth]{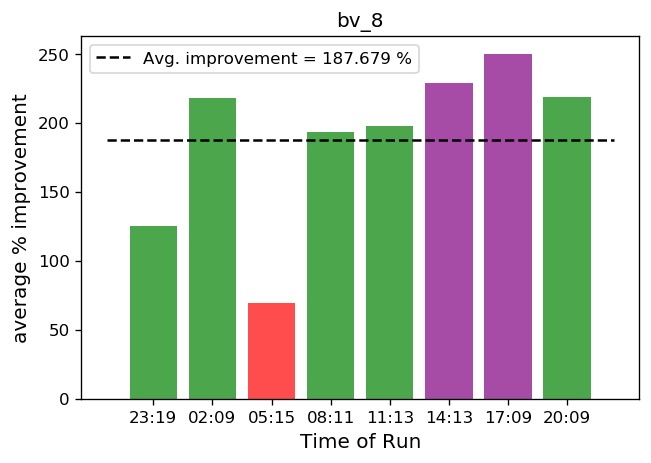}
  \end{subfigure}

  \begin{subfigure}{.3\textwidth}
    \includegraphics[width=\textwidth]{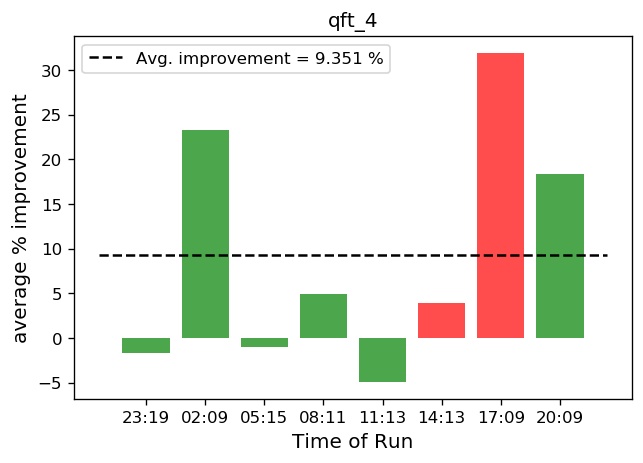}
  \end{subfigure}
  \hfil
  \begin{subfigure}{.3\textwidth}
    \includegraphics[width=\textwidth]{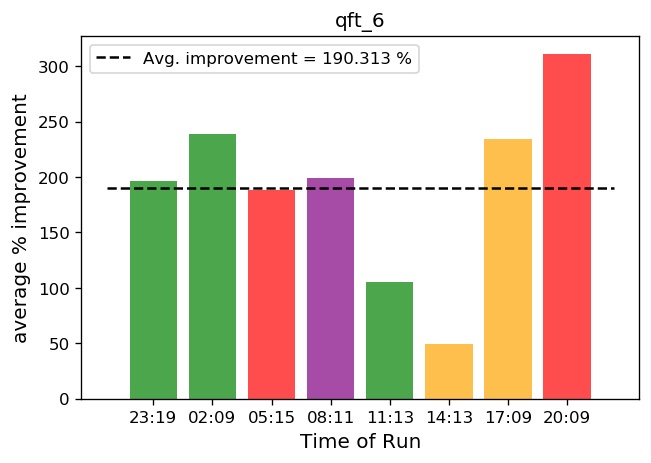}
  \end{subfigure}
  \hfil
  \begin{subfigure}{.3\textwidth}
    \includegraphics[width=\textwidth]{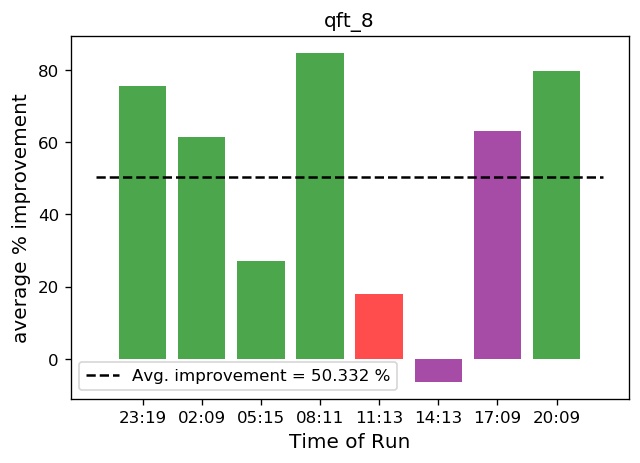}
  \end{subfigure}

  \begin{subfigure}{.3\textwidth}
    \includegraphics[width=\textwidth]{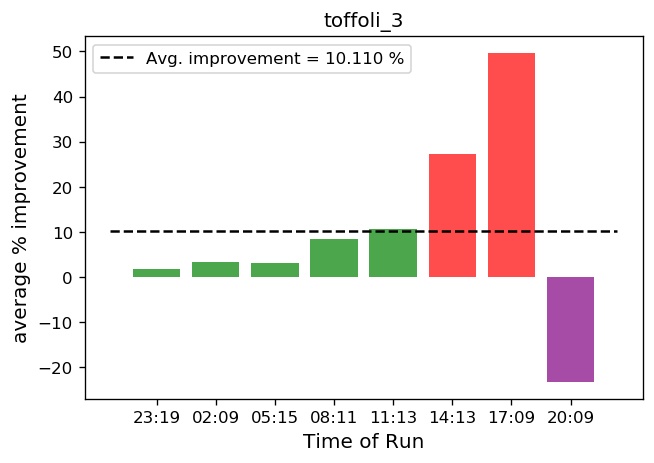}
  \end{subfigure}
  \hfil
  \begin{subfigure}{.3\textwidth}
    \includegraphics[width=\textwidth]{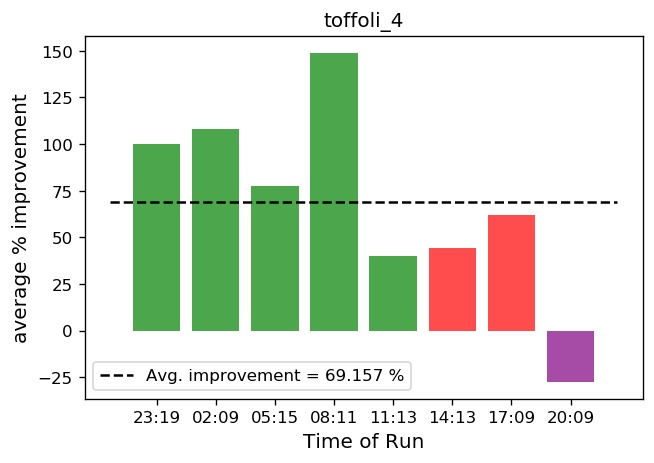}
  \end{subfigure}

  \begin{subfigure}{.3\textwidth}
    \includegraphics[width=\textwidth]{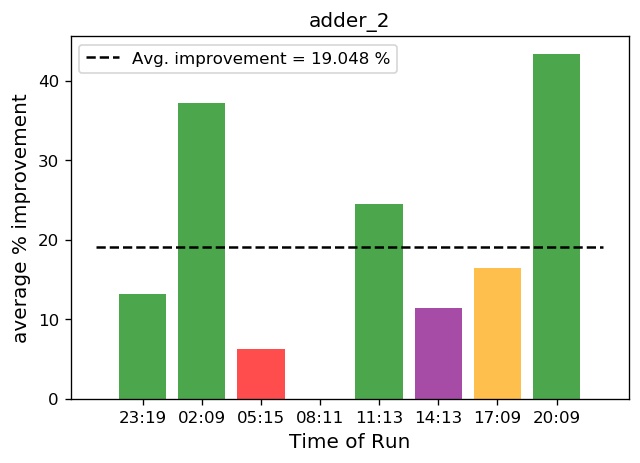}
  \end{subfigure}
  \hfil
  \begin{subfigure}{.3\textwidth}
    \includegraphics[width=\textwidth]{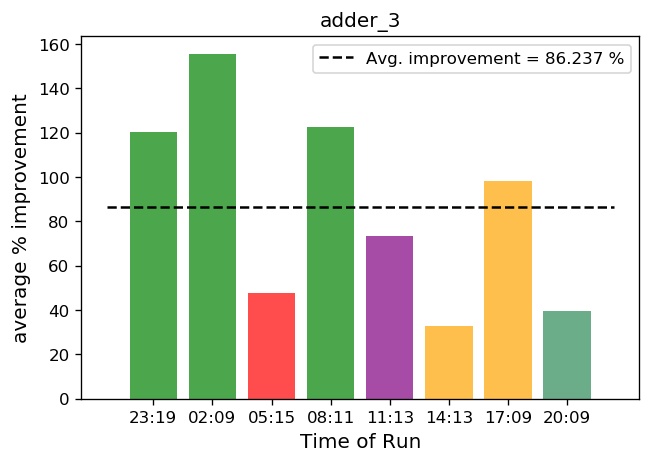}
  \end{subfigure}
  \hfil
  \begin{subfigure}{.3\textwidth}
    \includegraphics[width=\textwidth]{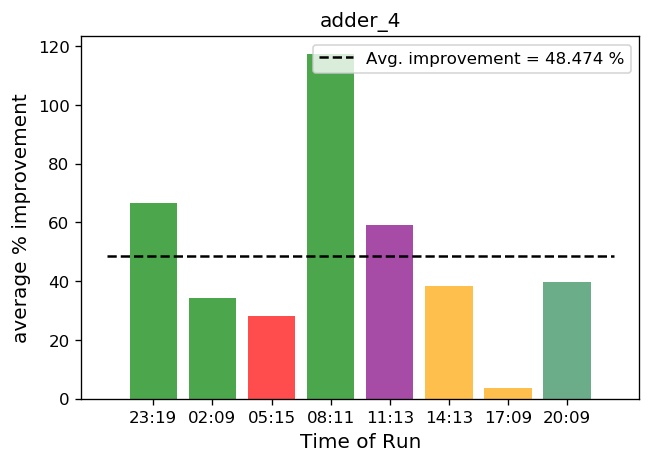}
  \end{subfigure}

  \caption{Percent improvement of accuracy for just-in-time
    transpilation in dedicated mode.}
  \label{fig:improvement}
\end{figure*}

\textbf{Observation 4:} Just-in-time transpilation offers more 
significant benefits when error data is obtained immediately prior to an
application circuit, irrespective of circuit width and depth.

We also conducted a sequence of experiments in a single 1-hour slot,
where the IBM-transpiled benchmark was run back-to-back with four
instances of (a) re-calibration (obtaining fresh error data) used by
just-in-time transpilation followed by (b) executing all
benchmarks. Our method was superior in all cases except for
adder(4), qft(6), toffoli(3), and sometimes better/sometimes worse
for qft(8).

\textbf{Observation 5:} Even when error data is obtained immediately
prior to an application run, just-in-time transpilation cannot {\em
  always} guarantee to provide superior results. Variations are more
pronounced long-term but also exist to a smaller extent
short-term. Best layouts change even within minutes.

\subsection{Detailed Accuracy Improvement for Dedicated Mode}

Figure~\ref{fig:improvement} depicts the average percent improvement
in accuracy for dedicated benchmark runs on the IBM Paris device
normalized to just-in-time transpilation with IBM's transpilation as a
baseline. Each bar corresponds to a separate run in a dedicated time
slot over a 24-hour period, i.e., 8 time slots in total.  Different
colors indicate different mappings.

Overall, most cases show a moderate to significant improvement
with the occasional exception of an insignificant loss (a few
instances of qft(4) and qft(8)) and few more significant losses (one
instance each for hs(6), hs(8), toffoli(3), toffoli(4)).
In terms of absolute accuracy, there was one data point for hs(6)
where IBM's result (59\%) was better than ours (52\%), and another in
hs(8) with 52\% vs. 44\% within the same benchmark run. We do not have
an explanation as neither hs(4) nor any other benchmark in the same
run showed inferiority of our method. The same holds for the last run
for toffoli(3) and toffoli(4). All these outliers have in common that
they use a never-seen-before layout, which may indicate that the error
collection method could possibly be improved on.

The overall average in improvement (over all 8 runs) is indicated by a
dashed line.

\textbf{Observation 6:} Just-in-time transpilation tends to improve
the relative accuracy of measured results on average by 3\%-190\% and up to
150\% in extreme cases in dedicated mode. Best layouts change within minutes.

In summary, Figure~\ref{fig:improvement} reinforces the last two
observations in that just-in-time transpilation provides benefits in
the majority of cases, but there are exceptions.

\subsection{Circuit Layout Analysis}

\begin{figure*}
  \centering

  \begin{subfigure}{.49\textwidth}
    \includegraphics[width=\textwidth]{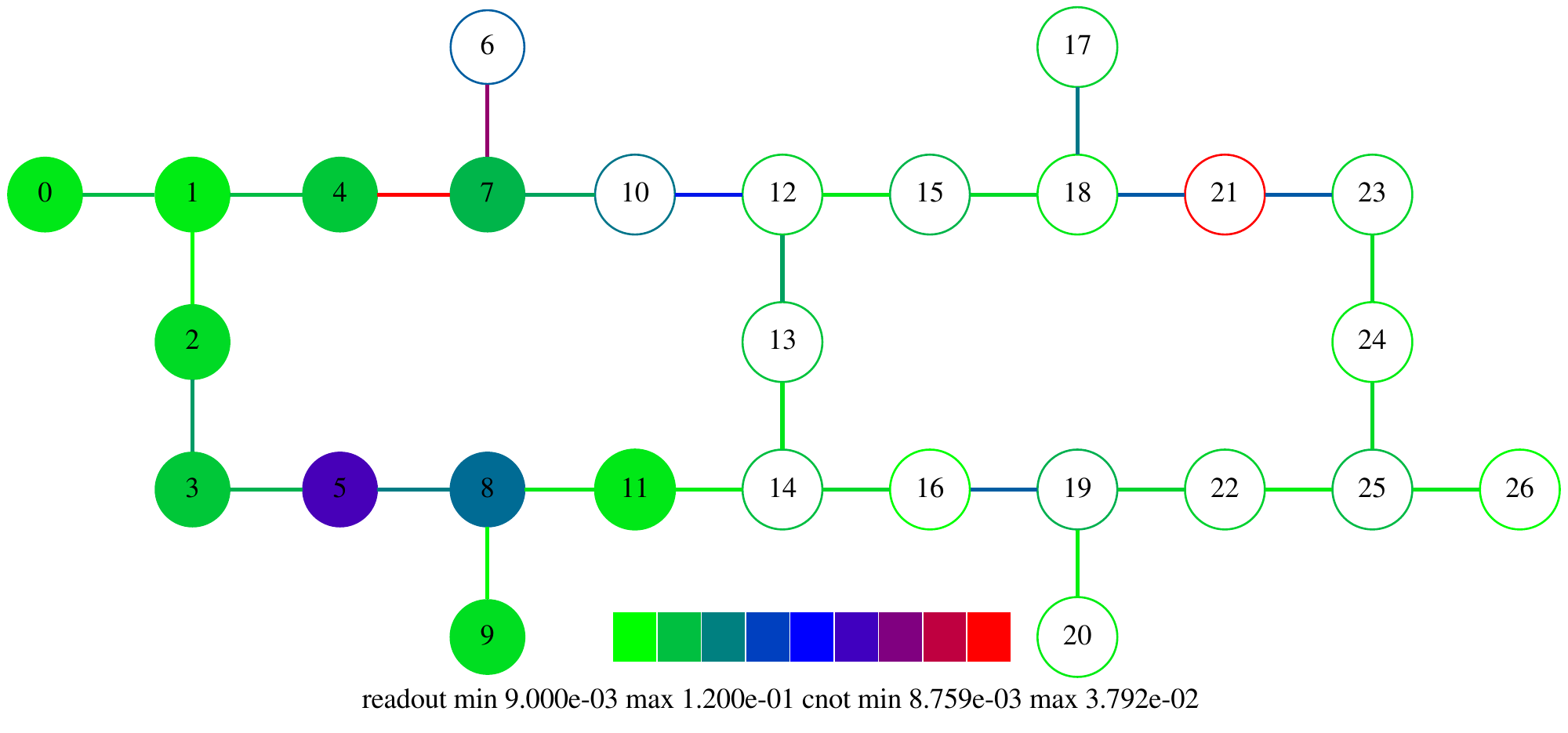}
    \caption{IBM layout for adder(4) on Paris}
    \label{fig:ibm-adder4-layout}
  \end{subfigure}
  \hfil
  \begin{subfigure}{.49\textwidth}
    \includegraphics[width=\textwidth]{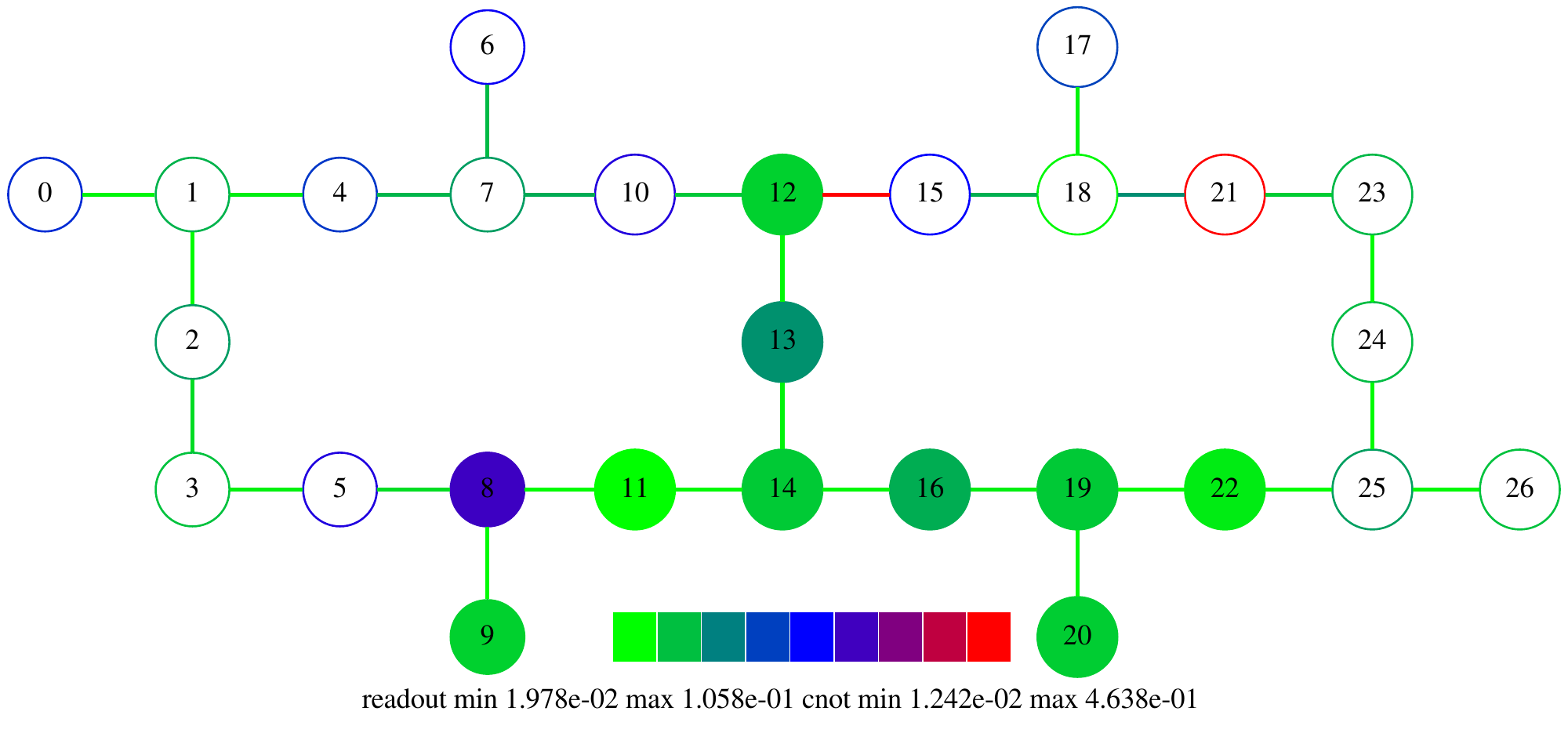}
    \caption{Our layout for adder(4) on Paris}
    \label{fig:our-adder4-layout}
  \end{subfigure}

  \begin{subfigure}{.49\textwidth}
    \includegraphics[width=\textwidth]{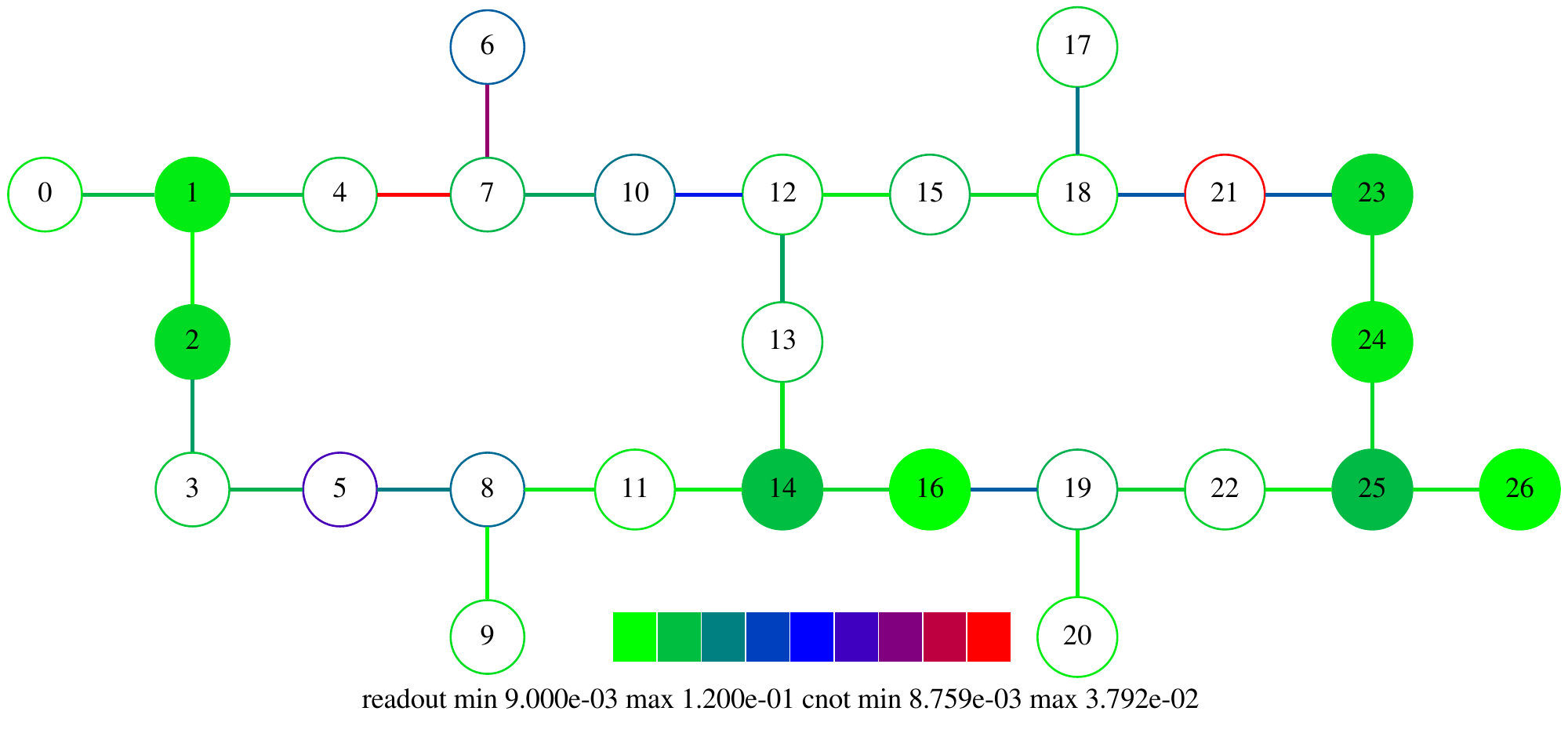}
    \caption{IBM layout for hs(8) on Paris}
    \label{fig:ibm-hs8-layout}
  \end{subfigure}
  \hfil
  \begin{subfigure}{.49\textwidth}
    \includegraphics[width=\textwidth]{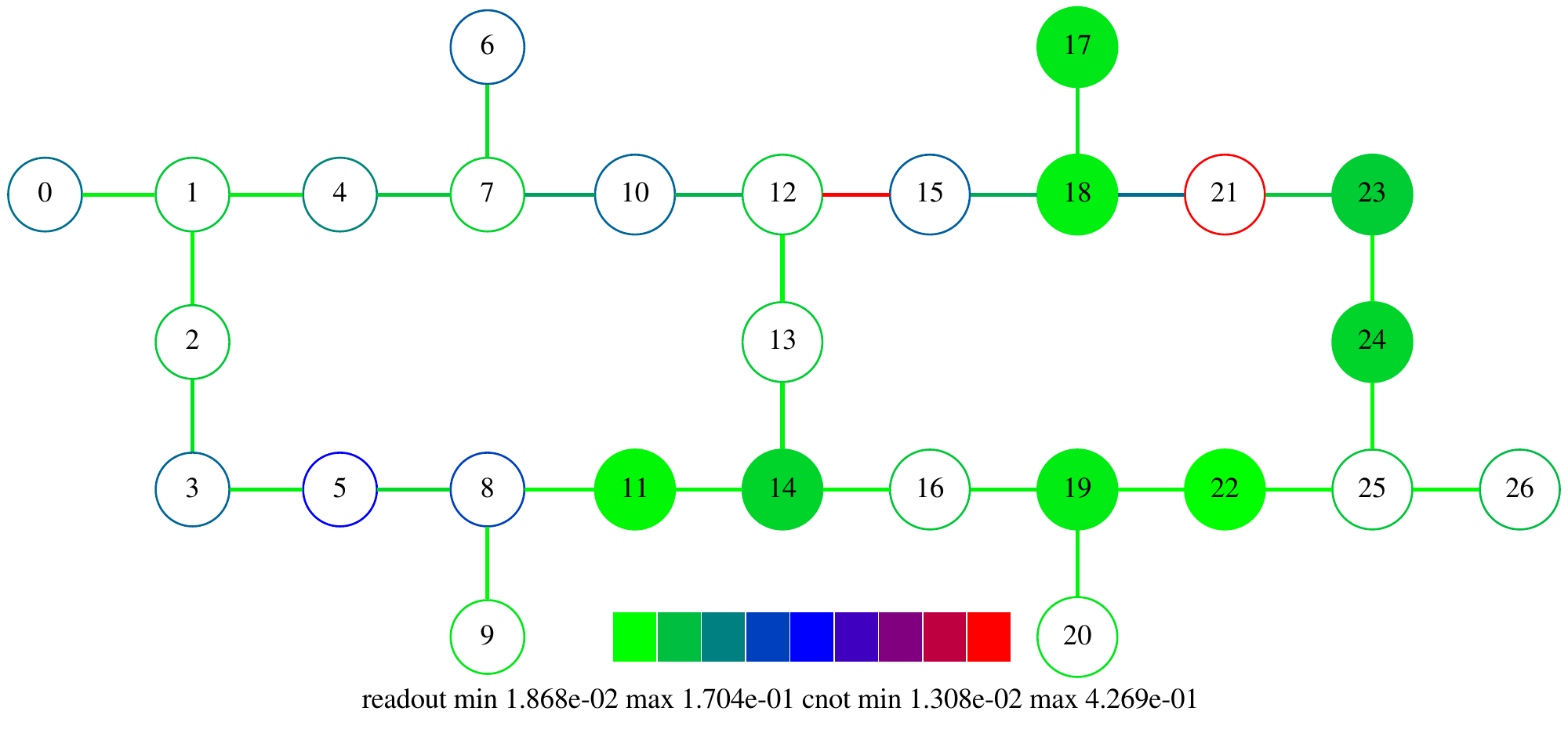}
    \caption{Our layout for hs(8) on Paris}
    \label{fig:our-hs8-layout}
  \end{subfigure}

\caption{Circuit Layouts}
\label{fig:layouts}
\vspace*{-\baselineskip}
\end{figure*}

Results so far have shown that differences in accuracy are correlated
to just-in-time transpilation on recent error data. We investigated
the benefits in a sensitivity study by considering changes in
virtual-to-physical qubit mappings. To this end, the resulting virtual
layouts were superimposed on the heatmap-coded interconnect of a
quantum device. Figure~\ref{fig:layouts} depicts pairs of IBM/our
layouts for hs(8) and adder(4). The nodes are qubits and edges are
couplings. A solidly colored qubit indicates that this qubit is used
within the respective circuit. Heatmaps range from low errors (green)
over blue to high errors (red) on a scale indicated for each graph,
i.e., separately for per-qubit readouts and couplings.

Overall, we can compare the errors of the IBM model (left) with that
of our error data (right) agnostic of any circuit. The error values
differ for a number of qubits and couplings, most notably couplings
4-7, 6-7, 5-8, and 12-15, and also qubits 0, 4, 5, 8, 15 and 17.  Others are
constantly good (many qubits and couplings remain green on both sides)
or constantly bad (e.g., qubit 21).

In the adder(4) example, our layout provides worse accuracy than
IBM's. First, we observe that in Figure~\ref{fig:ibm-adder4-layout}
coupling 4-7 within the circuit has high errors (red), and qubits 5
and 8 have mediocre fidelity (blue/purple).  In contrast, all couplings
in Figure~\ref{fig:our-hs8-layout} are of higher fidelity (green)
while only qubit 8 has lower fidelity (purple). Yet, IBM's accuracy at
10\% is better than ours at 8\%. Closer inspection reveals that our
lower end of the error spectrum has twice the error value of IBM's
lower end errors for both readouts and connectors. This means the
color spectrum on the right side should be shifted toward
higher errors. Another significant difference is in the readout
qubits, which are 4,7,8,9,11 for IBM's and 8,12,13,20,22 for our
transpiled code. This accounts in part of the difference in accuracy,
as will be discussed in the next subsection.

In the hs(8) example, our layout provides better accuracy than
IBM's. We observe that the selected qubits and couplings for the
circuit appear nearly equally good in Figure~\ref{fig:ibm-hs8-layout}
and Figure~\ref{fig:our-hs8-layout}, with a slight bias to higher
fidelity (lighter green) on the right side for qubits. As all qubits are
read out, this could explain the difference, even after taking into
account the differences in heatmap encoding. Notice that the hidden
shift algorithm requires only pairs of two qubits to be coupled, which
explains the layouts of isolated qubit pairs.

\textbf{Observation 7:} Differences in layouts corroborate the
hypothesis that there are two classes of errors: ``Persistent'' errors due to
low fidelity qubits and couplings that retain high errors, and
``transient'' errors that vary over shorter times. However, detailed analysis
of layouts with respect to noise levels of qubit readouts and
connectors remain only partially conclusive.

\subsection{Discussion}

The detailed analysis of layouts did not provide the clarity on a
case-by-case basis that we had anticipated. It is possible that other
factors have to be accounted for to explain differences in accuracy. 
In particular, it would be important to compare IBM's codes for determining error
rates with ours as we see much higher rates. This could be due to the
fact that the last calibration occurred hours ago, or it could
indicate that our algorithms are more suitable to find good layouts.
Furthermore, cross talk error is known to be in the order of readout
and coupling errors. Single qubit gate errors are said to be an order
of magnitude lower, as also reported by IBM after each
calibration. Another factor is the number of times a coupling is used
in conditional gates (e.g., CNOT) and, to a lesser extent, the number
of single qubit gates.
While we saw ``permanently'' high qubit and coupling errors for a few
device elements, most of them either remain at higher fidelity or
change in the medium range over time. It may be possible to further
distinguish errors within time ranges of minutes vs. hours, but we do
not have sufficient data to reliably do so.

With the results shown above, we conclude that dynamic on-the-fly
error calibration helps in taking into account the current state of
the qubits. Transpiling the circuits just-in-time with this error
information statistically produces more accurate results than those
produced when the IBM's published error information is used to
transpile the same circuits. 

This work has the following implications:
Our approach can help in producing better results in circuits from a
statistical perspective, but does not eliminate errors.

{\bf Recommendation 1:} We suggest to first obtain fresh error data
from a device before running sensitive circuits.

This is easily done in IBM's dedicated mode but even provides benefits
when hours lie in between obtaining error data and running the
just-in-time transpiled circuit.
An ensemble of circuits could then be prepared by
transpiling with the dynamically measured readout error information,
measured CNOT information, or both.

{\bf Recommendation 2:} A circuit transpiled with the default error
data should be included in experiments.

Sometimes, IBM's layout is superior, and a diversity of mappings can
provide more accurate results~\cite{DBLP:conf/micro/TannuQ19}.

{\bf Recommendation 3:} Devices should either be recalibrated more
frequently, or their errors should be assessed more often
(possibly both), with results made accessible to users.

If users always prefaced their code with a fresh error data
analysis, less science could be performed on a quantum device, yet
results may be of higher value. This is a subtle conundrum, and the
frequency of recalibration should be revisited by quantum backend operators.

Currently, the device properties (\textit{backend.properties()}
object) contain the calibration information published by IBM earlier
during the day. We suggest that IBM also publish more dynamic error
information along with the accuracy of the circuits by periodically
running these error extraction circuits. The developers could then,
based on the accuracy of results, decide whether or not to exploit
this dynamic error information to transpile their circuits or call
their own error measurement jobs. Furthermore, noise-based
transpilation (level 3 in Qiskit) should be the default.  Finally, job
dependencies and server-side transpilation should be introduced in
fairshare user mode to allow a second job to be transpiled depending
on output data of the first job that ran minutes before.

\section{Related Work}
\label{sec:related}

Current NISQ machines require substantial tuning of control signals in
order to compensate for noise in individual devices.  The closest
related work to ours focused on noise-aware mappings and read-out
errors~\cite{murali2019noise}, which is using noise data to adapt
qubit mappings during the transpilation process. This technique was
later integrated into IBM's Qiskit transpilation, which uses daily
calibrations for qubit mappings. As our work shows, more frequent
noise recalibrations provide additional benefits on today's NISQ
devices.

Other techniques focus on interpreting different qubit mappings
statistically and inverting computational results to benefit from
lower errors in non-excited
states~\cite{DBLP:conf/micro/TannuQ19,DBLP:conf/micro/TannuQ19a},
hardware-specific optimizations confined to back-end passes of the
compiler across different NISQ platforms~\cite{Murali2019}, or
reduction in cross talk~\cite{murali20}.  IBM uses
pulses to further reduce noise~\cite{ibm-pulses}, a technique that was
generalized to larger circuits or blocks of gates with shorter
pulses~\cite{chong-asplos19,gokhale19}.

Other qubit mapping approaches were shown to be effective for
smaller-scale NISQ
devices~\cite{murali2019noise,zulehner2018efficient,Tannu:2019:QCE:3297858.3304007}
but often required high time/memory consumption when scaling up, while
others had more scalable algorithms but compromised in the fidelity of
the mapping~\cite{wille2016look,siraichi2018qubit}, while yet others
focused on scalability without considering noise details at the
same level of detail~\cite{li2019tackling}, or used dynamic
assertions as a means to filter by noise~\cite{liu20}.
These techniques can orthogonally improve results on top of our
recalibration.

\section{Conclusion}
\label{sec:conc}

We have contributed a methodology for on-the-fly transpilation taking
fresh error data for readouts and two-qubit gates into
account. Our experiments have shown the effectiveness of this
technique on current NISQ devices resulting in 3-190\% improvement of
accuracy for dedicated execution and 8-304\% for shared job queues
with a maximum observed improvement of a factor of four,
depending on the circuit. Improvements are best when error data was
recently obtained, leading to recommendations for adjusting operations
of quantum devices to obtain and publish error data more frequently.

\bibliographystyle{IEEEtran}
\bibliography{mybib,martonosi,chong,brown,qureshi,yufei}

\begin{thebibliography}{10}
\providecommand{\url}[1]{#1}
\csname url@samestyle\endcsname
\providecommand{\newblock}{\relax}
\providecommand{\bibinfo}[2]{#2}
\providecommand{\BIBentrySTDinterwordspacing}{\spaceskip=0pt\relax}
\providecommand{\BIBentryALTinterwordstretchfactor}{4}
\providecommand{\BIBentryALTinterwordspacing}{\spaceskip=\fontdimen2\font plus
\BIBentryALTinterwordstretchfactor\fontdimen3\font minus
  \fontdimen4\font\relax}
\providecommand{\BIBforeignlanguage}[2]{{%
\expandafter\ifx\csname l@#1\endcsname\relax
\typeout{** WARNING: IEEEtran.bst: No hyphenation pattern has been}%
\typeout{** loaded for the language `#1'. Using the pattern for}%
\typeout{** the default language instead.}%
\else
\language=\csname l@#1\endcsname
\fi
#2}}
\providecommand{\BIBdecl}{\relax}
\BIBdecl

\bibitem{murali2019noise}
P.~Murali, J.~M. Baker, A.~J. Abhari, F.~T. Chong, and M.~Martonosi,
  ``Noise-adaptive compiler mappings for noisy intermediate-scale quantum
  computers,'' \emph{arXiv preprint arXiv:1901.11054}, 2019.

\bibitem{Tannu:2019:QCE:3297858.3304007}
S.~S. Tannu and M.~K. Qureshi, ``Not all qubits are created equal: A case for
  variability-aware policies for {NISQ}-era quantum computers,'' in
  \emph{Proceedings of the Twenty-Fourth International Conference on
  Architectural Support for Programming Languages and Operating Systems}, 2019.

\bibitem{Qiskit}
\BIBentryALTinterwordspacing
H.~Abraham, AduOffei, I.~Y. Akhalwaya, and et~al., ``Qiskit: An open-source
  framework for quantum computing,'' 2019. [Online]. Available:
  \url{qiskit.org}
\BIBentrySTDinterwordspacing

\bibitem{DBLP:conf/micro/TannuQ19}
S.~S. Tannu and M.~K. Qureshi, ``Ensemble of diverse mappings: Improving
  reliability of quantum computers by orchestrating dissimilar mistakes,'' in
  \emph{International Symposium on Microarchitecture}, 2019, pp. 253--265.

\bibitem{DBLP:conf/micro/TannuQ19a}
------, ``Mitigating measurement errors in quantum computers by exploiting
  state-dependent bias,'' in \emph{International Symposium on
  Microarchitecture}, 2019, pp. 279--290.

\bibitem{Magesan_2011}
\BIBentryALTinterwordspacing
E.~Magesan, J.~M. Gambetta, and J.~Emerson, ``Scalable and robust randomized
  benchmarking of quantum processes,'' \emph{Physical Review Letters}, vol.
  106, no.~18, May 2011. [Online]. Available:
  \url{http://dx.doi.org/10.1103/PhysRevLett.106.180504}
\BIBentrySTDinterwordspacing

\bibitem{ibmqexp}
``{IBM Quantum Experience},'' \url{https://github.com/Qiskit/qiskit-api-py},
  accessed: 2018-11-16.

\bibitem{Fairshare}
\BIBentryALTinterwordspacing
IBM, ``Fair-share queuing,'' 2020. [Online]. Available:
  \url{https://quantum-computing.ibm.com/docs/cloud/backends/queue}
\BIBentrySTDinterwordspacing

\bibitem{Murali2019}
P.~Murali, N.~M. Linke, M.~Martonosi, A.~J. Abhari, N.~H. Nguyen, and C.~H.
  Alderete, ``Full-stack, real-system quantum computer studies: Architectural
  comparisons and design insights,'' in \emph{International Symposium on
  Computer Architecture}, 2019, pp. 527--540.

\bibitem{murali20}
\BIBentryALTinterwordspacing
P.~Murali, D.~C. Mckay, M.~Martonosi, and A.~Javadi-Abhari, ``Software
  mitigation of crosstalk on noisy intermediate-scale quantum computers,'' in
  \emph{Proceedings of the Twenty-Fifth International Conference on
  Architectural Support for Programming Languages and Operating Systems}, ser.
  ASPLOS ’20.\hskip 1em plus 0.5em minus 0.4em\relax New York, NY, USA:
  Association for Computing Machinery, 2020, p. 1001–1016. [Online].
  Available: \url{https://doi.org/10.1145/3373376.3378477}
\BIBentrySTDinterwordspacing

\bibitem{ibm-pulses}
L.~Bishop and J.~Gambetta, ``Reduction and/or mitigation of crosstalk in
  quantum bit gates,'' 2019, {US} Patent App. 15/721,194.

\bibitem{chong-asplos19}
\BIBentryALTinterwordspacing
Y.~Shi, N.~Leung, P.~Gokhale, Z.~Rossi, D.~I. Schuster, H.~Hoffmann, and F.~T.
  Chong, ``Optimized compilation of aggregated instructions for realistic
  quantum computers,'' in \emph{Proceedings of the Twenty-Fourth International
  Conference on Architectural Support for Programming Languages and Operating
  Systems}, ser. ASPLOS '19.\hskip 1em plus 0.5em minus 0.4em\relax New York,
  NY, USA: ACM, 2019, pp. 1031--1044. [Online]. Available:
  \url{http://doi.acm.org/10.1145/3297858.3304018}
\BIBentrySTDinterwordspacing

\bibitem{gokhale19}
P.~Gokhale, Y.~Ding, T.~Propson, C.~Winkler, N.~Leung, Y.~Shi, D.~I. Schuster,
  H.~Hoffmann, and F.~T. Chong, ``Partial compilation of variational algorithms
  for noisy intermediate-scale quantum machines,'' in \emph{International
  Symposium on Microarchitecture}, 2019.

\bibitem{zulehner2018efficient}
A.~Zulehner, A.~Paler, and R.~Wille, ``An efficient methodology for mapping
  quantum circuits to the {IBM} {QX} architectures,'' \emph{IEEE Transactions
  on Computer-Aided Design of Integrated Circuits and Systems}, 2018.

\bibitem{wille2016look}
R.~Wille, O.~Keszocze, M.~Walter, P.~Rohrs, A.~Chattopadhyay, and R.~Drechsler,
  ``Look-ahead schemes for nearest neighbor optimization of {1D} and {2D}
  quantum circuits,'' in \emph{2016 21st Asia and South Pacific Design
  Automation Conference (ASP-DAC)}.\hskip 1em plus 0.5em minus 0.4em\relax
  IEEE, 2016, pp. 292--297.

\bibitem{siraichi2018qubit}
M.~Y. Siraichi, V.~F.~d. Santos, S.~Collange, and F.~M.~Q. Pereira, ``Qubit
  allocation,'' in \emph{Proceedings of the 2018 International Symposium on
  Code Generation and Optimization}.\hskip 1em plus 0.5em minus 0.4em\relax
  ACM, 2018, pp. 113--125.

\bibitem{li2019tackling}
\BIBentryALTinterwordspacing
G.~Li, Y.~Ding, and Y.~Xie, ``Tackling the qubit mapping problem for {NISQ}-era
  quantum devices,'' in \emph{Proceedings of the Twenty-Fourth International
  Conference on Architectural Support for Programming Languages and Operating
  Systems}, ser. ASPLOS '19.\hskip 1em plus 0.5em minus 0.4em\relax New York,
  NY, USA: ACM, 2019, pp. 1001--1014. [Online]. Available:
  \url{http://doi.acm.org/10.1145/3297858.3304023}
\BIBentrySTDinterwordspacing

\bibitem{liu20}
\BIBentryALTinterwordspacing
J.~Liu, G.~T. Byrd, and H.~Zhou, ``Quantum circuits for dynamic runtime
  assertions in quantum computation,'' in \emph{Proceedings of the Twenty-Fifth
  International Conference on Architectural Support for Programming Languages
  and Operating Systems}, ser. ASPLOS ’20.\hskip 1em plus 0.5em minus
  0.4em\relax New York, NY, USA: Association for Computing Machinery, 2020, p.
  1017–1030. [Online]. Available:
  \url{https://doi.org/10.1145/3373376.3378488}
\BIBentrySTDinterwordspacing

\end{thebibliography}

\end{document}